\documentstyle[preprint,eqsecnum,aps]{revtex}
\begin{document}
\draft
\preprint{BA-93-53}
\title{
A Simple Model of Large Scale Structure Formation
}
\author{ R.~K.~Schaefer and Q.~Shafi  }
\address{
 Bartol Research Institute \\
 University of Delaware, Newark, DE 19716\\
}
\date{\today}
\maketitle

\begin{abstract}
We explore constraints on inflationary models employing data on large scale
structure mainly from COBE temperature anisotropies and IRAS selected galaxy
surveys, taking care not to apply linear perturbation theory to data in the
non-linear regime.  In models where the tensor contribution to the COBE signal
is negligible, we find the spectral index of density fluctuations $n$ must
exceed 0.7.  Furthermore the COBE signal cannot be dominated by the tensor
component, implying $n>0.85$ in such models.  The data favors cold plus hot
dark matter models with $n$  close to unity and $\Omega_{HDM} \sim 0.20 -
0.35$.  We present realistic grand unified  theories, including supersymmetric
versions,  which produce inflation with these properties.
\end{abstract}

\narrowtext

\section{Introduction}
\label{sec:level1}

   The inflationary universe scenario \cite{inflrev} provides an attractive
resolution of the well known ``horizon", ``flatness", and ``causality" puzzles
encountered in the standard big bang cosmology.  This scenario is most easily
realized within the framework of both ordinary and supersymmetric grand unified
theories (GUTS).   In the simplest realizations of inflation, the density
fluctuations are Gaussian and close to the Harrison-Zeldovich form, and the
background density equals the critical density $\rho_c$.  Primordial
nucleosynthesis implies that less than 10\% of the background density is
composed of baryons, and so the bulk of matter is ``dark", presumably in the
form of relic elementary particles.

   Almost a decade ago it was pointed out by Shafi and Stecker \cite{ss84} that
two types of dark matter, cold and hot, would provide, within the inflationary
context, an elegant way of understanding observations which indicated a
surprising amount of clustering on larger than cluster scales.   Since only the
cold component clustered on smaller scales, and the hot component only clusters
on larger scales, such a universe would have an  enhanced large scale
clustering power.  Examples of GUTS containing cold  plus hot dark matter
(C+HDM) were also presented.  The implications of this  picture for microwave
background anisotropies were worked out in 1989  \cite{sss,holtzman89}.  Mass
functions \cite{occhionero88}  bulk streaming motions  \cite{sss,holtzman89},
cluster number densities and correlation functions, as well as ``great
attractors" were also found to be compatible \cite{vdns92,holtzman93} with
data.

   Recent data on large scale structure from a variety of sources, particularly
COBE and IRAS selected galaxy surveys, provide additional strong support to
this remarkably ``simple" scenario for structure formation.  The COBE group
\cite{smoot92} found an anisotropy  amplitude which they characterized
by an averaged quadrupole moment  amplitude  $\delta T/T = Q_{rms-PS}/T = 6.2
\pm 1.5\times 10^{-6}$.  As we have emphasized \cite{ssnature}, the earlier
predictions \cite{sss,holtzman89} for C+HDM models  (3/4 CDM, 1/4 HDM) with
negligible baryonic matter content were $7.8\times 10^{-6}/b_8$, while  for CDM
alone they were significantly smaller ($4.7\times 10^{-6}/b$), where  $b_8 =
$(rms mass fluctuation on the scale $8\ h^{-1}$Mpc)$^{-1}$.   Including a  5\%
baryon density (implied by primordial nucleosynthesis) modifies these
predictions to $8.2\times 10^{-6}/b_8$ for this C+HDM mixture and  $5.1\times
10^{-6}/b_8$  for CDM.  Observationally derived values of $b_8$ seem to fall in
the  range $1.3<b_8<2.5$, meaning the COBE result was in remarkable agreement
with C+HDM models.  COBE also lent some weak support for the inflationary power
spectrum; the data are consistent with the $P(k)\propto k^n$ spectrum with
$n=1.1\pm 0.5$ \cite{smoot92}.  Following the COBE result other
pieces of evidence pointing to the C+HDM model were discovered: compatibility
with early quasar formation  \cite{ss93,haehnelt93}, compatibility with
galactic correlations and  velocities, \cite{davis92,klypin93,jing93}, APM
correlations  \cite{jing93,efstatbw} quiet local Hubble flow \cite{schlegel93},
``cosmic  mach number" \cite{xiang92}, and counts in cells
\cite{taylor92,pogosyan93}.   Thus, there seems no doubt that the C+HDM model
is a candidate worthy of  serious consideration.

    While we know that the primordial density power spectrum $P(k)\propto k$
with C+HDM provides a good fit to the data, it has long been known that the
inflationary scenario does not quite yield this simple form.  In recent
studies, these correction factors have been exploited to yield  power spectra
$P(k)\propto k^n$ with $n$ significantly less than 1 as a  means to repair the
relative large/small scale problems of the pure CDM  model.  In this paper, one
of our tasks will be to undertake a study of the range of allowed $n$.  [Note
that in this work we will only consider
$n {\mathrel{\raise.3ex\hbox{$<$\kern-.75em\lower1ex\hbox{$\sim$}}}}1$,
although inflationary models where $n$ exceeds unity can be constructed.]
Since
we do not precisely know the HDM fraction of the universe, we will explore the
two  dimensional parameter space $n-\Omega_{HDM}$, and examine the constraints
imposed by the data.

   We do this by comparing the model predictions with data from COBE, the
power spectrum of IRAS galaxies, IRAS counts-in cells, and bulk velocities
from the POTENT analysis.  We will also consider requirements for the early
formation of quasars.  After fitting the normalization and the bias  factor for
IRAS galaxies, we calculate $\chi^2$ for each model.  We then  compare the
results of the models and present $\chi^2$ contour plots in the  $n -
\Omega_{HDM}$ fraction plane.  We also consider the possibility that  some of
the COBE signal was produced by the long wavelength gravity waves generated
during inflation.

 Because of uncertainties in the simulation of  structure formation in the
non-linear gravity regime, we will concentrate  chiefly on structure which is
still described by linear perturbation theory  (scales $>20 h^{-1}\ Mpc$).
This avoids  the difficulties inherent in identifying galaxies and clusters and
separating out complicated dynamic effects, {\it e.g.}, the velocity bias of
galaxies.  Our main aim is to explore a large region of parameter space, and
try to identify that part of it for  which running more detailed computer
simulations would make the most sense.

   Finally, if we are to take seriously the whole picture of inflation and
C+HDM we need to identify plausible models of inflation which are compatible
with our analysis of the large scale data.  Here we will  consider only
``realistic" models of inflation.  By this, we mean models  in which the
inflating field is a part of a larger theory which contains the  following
elements:
\begin{itemize}

\item neutrinos with masses in the eV range.

\item a cold dark matter candidate.

\item a high energy particle physics structure which is compatible with low
energy physics and its hints about higher symmetries.

\item a successful baryogenesis following inflation.
\end{itemize}
Without these elements we find the inflation model to be somewhat {\it ad hoc}.

    The plan of the paper is as follows.  In the next section, we will
describe our testing of inflationary models with large scale structure  data in
the linear perturbation regime using a $\chi^2$ analysis.  This will turn out
to strongly limit the power spectrum.   In section III we will consider some
general constraints from data which come from structure in the non-linear
regime.  This will limit mainly the HDM fraction.  In section IV we consider
models of inflation which satisfy all of  our requirements  for ``realistic"
inflation models.  In particular we present some new examples of chaotic
inflation based on supersymmetric grand unification.  We end with some general
conclusions in section V.  We have attempted to make this paper somewhat
self-contained for a more general audience.

\section{Constraints from Large Scale Structure Data in the Linear Regime.}
\label{sec:level2}

     In this section, we will describe data and our $\chi^2$ analysis of
theoretical predictions for these observations, but first we will discuss a
few general issues concerning the models and the testing strategy employed
here.

    First of all, the problem with drawing strong conclusions about any given
model of structure formation is that there are so many parameters to vary that
we have a multi-dimensional parameter space to explore.  We will choose to take
best guess values of three of them, the baryonic fraction $\Omega_{baryon} =
\rho_{baryon}/\rho_c$, the cosmological constant $\Lambda$, and the  Hubble
constant, $H_0 = 100 h$ km sec$^{-1}$,  with $h$ observationally constrained to
the values
$0.4 {\mathrel{\raise.3ex\hbox{$<$\kern-.75em\lower1ex\hbox{$\sim$}}}} h
{\mathrel{\raise.3ex\hbox{$<$\kern-.75em\lower1ex\hbox{$\sim$}}}} 1.0$.
Constraints on the age of the universe
from globular star clusters, nuclear  cosmochronology, and white dwarf ages
imply that high values of the Hubble  constant are forbidden, {\it i.e.}
$h<0.6$ in an  $\Omega=1$ universe.  Thus  these models will be allowed only if
the observations eventually settle into a more restricted
range $0.4<h<0.6$.  Since many  quantities vary as $h^2$ there is
still some freedom despite this narrow range of Hubble constant.  In the
present analysis we will use only the central value, $h=0.5$.

    Primordial nucleosynthesis strongly constrains the baryon density
\cite{walker} as $0.010\leq \Omega_{baryon} h^2 \leq 0.015$.  Using $h=0.5$,
we can express this 95\% confidence constraint as
\begin{equation}
 \Omega_{baryon} = 0.05 \pm 0.01
\end{equation}
We note that allowing for the uncertainty in $h$, the baryonic fraction
could range from $\Omega_{baryon} \sim 0.03 - 0.9$.
We intend to explore this uncertainty in the Hubble constant and baryon
fraction and its implications in a future publication.
Once we have fixed the baryon density, the other densities can be completely
specified by the hot dark matter density $\Omega_{HDM}$ as
$$\Omega_{CDM} = 1 - \Omega_{baryon} - \Omega_{HDM}.  $$
The hot dark matter fraction (combined with the hubble constant) also
specifies the neutrino mass.  If we have one flavor of neutrino whose mass
is in the eV range, usually taken to be $\nu_\tau$, then
\begin{equation}
\Omega_{HDM} = \left( {m_{\nu_\tau}\over 23\ {\rm eV}}\right) \left( {0.5\over
h}\right)^2.
\end{equation}

   We will set the cosmological constant ($\Lambda$) to zero.
There is evidence to support this choice.  The local (within 60
$h^{-1}$ Mpc) velocity field implies values of $\Omega$ close to unity
\cite{dekel93}.  Furthermore, in a $\Lambda$ dominated universe, there seems
to be too few gravitational quasar
lensing events, and the bulk streaming velocities are too small.

    The growth of density fluctuations is affected by the dynamics of
the matter content, producing a scale dependent modification of the density
fluctuation power spectrum.  The relative growth as a function of scale is
discussed in Appendix A, and the results are summarized in Figure 1, where we
present the transfer functions in Fourier space as a function of Fourier
wavenumber $k=2\pi/\lambda$.  In figure 1, we see that increasing the neutrino
fraction decreases the amount of growth, and hence the amplitude,  on small
scales (large $k$).  We also note that there are some modifications of the
shape of the spectrum on quite large scales. For example the transfer function
with $\Omega_{HDM} = 0.3$ is slightly  steeper at $k\sim 0.07 h$ Mpc$^{-1}$
than that of either  $\Omega_{HDM} = 0.0$ (CDM) or $\Omega_{HDM} = 0.5$.

    In order to do our comparison in the most unambiguous way, we will confine
our attention mostly  to the regime where linear theory is appropriate.  In the
past, it was common to normalize linear power spectra by use of the fact that
the rms  optically selected galaxy density fluctuations are $\delta N/N = 1$ on
a  scale of 8 $h^{-1}$ Mpc.   In the most naive case, where one assumes that
galaxies (light) trace the mass distribution, one would set the rms mass
fluctuation also equal  to one on this scale. More recently, however, it has
become apparent that the  galaxies are more strongly clustered than the mass,
i.e., they are biased tracers of the mass.  The usual method for dealing with
this complication is to assume that there is a linear relation between the
optical galactic density  and mass fluctuations using a bias parameter $b$,
which is independent of scale  $b = (\delta N/N)/(\delta M/M)$.  The mass
fluctuation $\sigma(R)$ in a  sphere of radius $R$ is calculated in linear
theory by
\begin{equation}
\sigma^2(R) \equiv {1\over 2 \pi^2} \int_0^\infty dk\ k^2 P_{th}(k)
\left[ 3 {j_1(kR)\over kR}\right]^2,
\end{equation}
where $j_1(x)$ is the first spherical Bessel function and the term in brackets
is the Fourier transform of a sharp edged sphere of radius $R$.  Clearly,  when
a perturbation amplitude approaches unity, linear theory is no longer
appropriate.   Even with $b$ as large as $2$, this implies $\sigma(8\ h^{-1}$
Mpc$) = 0.5$, which is still quite non-linear.   Using the spherical collapse
approximation (see Appendix B) we estimate that for $\sigma=0.5$, we are
already highly contaminated by non-linear effects.   If we consider scales for
which $\sigma (R) \leq 0.4$ we estimate that the  non-linear corrections will
be ${\mathrel{\raise.3ex\hbox{$<$\kern-.75em\lower1ex\hbox{$\sim$}}}} 30\%$.

    A look at figure 1 shows that the
greatest difference between the models considered here occurs on the small
scales.  Limiting our range to $k \leq 0.3 h$ Mpc$^{-1}$ implies that our
testing will have a somewhat weakened ability to discriminate between models
with different HDM fractions.  In section III we will consider
some constraints from non-linear structures to help us pin down the dark matter
fraction.

    We will first give a brief description of the large scale structure data,
followed by a description of our calculations for different models.

\subsection{Large Scale Structure Data}

  Here we will discuss the particulars of the data we are using.  We explain
our reasons for choosing the data and method of interpretation of this data.

\subsubsection{COBE data}

    The large amplitude of the COBE measured
temperature fluctuations is characterized by the extrapolated quadrupole moment
$Q_{rms-PS}$.
This amplitude was a factor of 2-3 larger than predicted in the usual
CDM models fit to
galactic  structure, and so gave strong support for the C+HDM models
\cite{ssnature}.  However,
various authors seem not content to use the COBE analysis of the amplitude.
Instead they choose to use the sky variance at $10^\circ$ which, when used
with a Gaussian beam, implies a somewhat smaller amplitude for the
fluctuations ({\it e. g.,} \cite{efstatbw,liddle93}
corresponding to $Q_{rms-PS} = 15.3\mu$K).  The COBE beam pattern, however,
is only approximately a Gaussian shape.  More careful analysis of the
COBE results \cite{wright93} using the actual beam pattern seems to
confirm the original higher value of the amplitude.  In addition, use of the
correct beam pattern reduces the variance of the fit amplitude, so the best
fit $n=1$ COBE amplitude now corresponds to $Q_{rms-PS} = 17.1\pm 2.9\ \mu$K.
 Although it
is customary to quote the results in terms of the quadrupole moment, this
corresponds to only the very small wavenumber end of the spectrum.  The best
fit
amplitude of the quadrupole moment will be somewhat dependent on the value of
$n$ used in the analysis.  In order to find a better quantity than the
quadrupole, Wright {\it et al.} \cite{wright93} recommend normalizing the
amplitude
to the
hexadecupole $\delta T_4 = 12.8\pm 2.3\ \mu$K, when $n\ne 1$, as in
the inflationary models we consider here.  This yields a best fit amplitude
less dependent on the value of $n$.

   In passing, we note that the COBE results should be taken seriously, as a
balloon experiment from the MIT/GSFC/Princeton collaboration
\cite{ganga93} sees the same temperature correlation function as COBE
and find similar values for the fluctuation
amplitude, which they specify with $Q_{rms-PS}$.
However, they currently only have data from the northern hemisphere and
their limits on $Q_{rms-PS}$ are not as constraining as those for COBE.  In the
future they plan to cover the southern hemisphere as well, which would improve
the limits of the COBE power spectrum exponent, which are currently
$n=1.1\pm0.5$.

  Ideally, we would like to include the detections of smaller scale anisotropy
experiments in this analysis.  There has been a wave of new detections of
temperature anisotropies on $1^\circ -2^\circ$ scales.  It is not clear that
all of the detections are giving a totally consistent picture, and even
different scans with the same instrument give different results.  A possible
explanation is that the systematic errors these experiments face are
quite complicated.  We expect that these uncertainties will be resolved
and that we will know the amount of degree scale anisotropy which exists with
some precision.  We also point out that C+HDM models, when normalized to COBE
produce the same degree scale anisotropies as similarly normalized CDM models.
 The only differences occur for anisotropy measurements $\ll 1^\circ$.  Since
the degree scale anisotropies are all in the right ballpark for $\Omega=1$,
$n\approx 1$ models normalized to COBE, we take this as an encouraging sign for
the models we consider in this paper.

\subsubsection{The IRAS Large Scale Survey of Galaxies}

The IRAS survey of galaxies done by the QDOT (Queen
Mary-Durham-Oxford-Toronto) collaboration \cite{efstathiou90},  extends as deep
as several hundred Mpc,  approaching the smallest scales observable by the COBE
satellite.   Combining the COBE data with the IRAS survey of galaxies thus
covers the  whole range of scales where the fluctuations can be described by
the linear theory.  The QDOT IRAS
survey has measured redshifts for 1 - in - 6 IRAS galactic sources (1824
galaxies) with  IRAS fluxes $>0.6$ Jy.  The IRAS selected galaxies seem to be
more uniformly distributed than optically selected galaxies and may therefore
give a fairer representation of the universe.  By concentrating on measuring
only 1 in 6 galaxies, the QDOT collaboration obtained a deep sparse sample out
to a depth larger than that of the Berkeley IRAS survey \cite{fisher92}
which measured
redshifts for all IRAS sources above a flux of 1.2 Jy.  Using the redshift of
the source as a distance indicator, combined with angular position data, one
has a three dimensional picture of the galaxy distribution.  This distribution
can be analyzed to directly extract the power spectrum of density
fluctuations \cite{fisher92,feldman93}.  The results of the
Feldman, et al. \cite{feldman93} analysis is shown as the power spectrum data
in Figure 2.

   We will use this QDOT IRAS data to test models with theoretical power
spectra given by
\begin{equation}
P_{th}(k) = A k^n [T(k)]^2.
\end{equation}
For $n\sim1$ and the transfer functions $T(k)$ in figure 1, we can see that the
power spectra go like $P(k) \sim k$ on very large scales and like $P(k)\sim
k^{-3} - k^{-4}$ on very small scales, with a peak somewhere around $k\sim
0.02-0.10$.  In order to get the proper normalization for the amount
of structure on scales up to the power spectrum peak scale, one must measure
the power on scales larger than the power spectrum peak scale.  In
the mass fluctuation integral (equation 2.3) there are significant
contributions to the mass fluctuations in 20 Mpc spheres coming from 100 Mpc
scales ($k\sim 0.03 h$/Mpc).  Thus we feel that to test power spectra of the
type considered here, we require a survey out to a depth at least as large as
the QDOT survey.  A comparison of the QDOT and 1.2 Jy Berkeley power spectra,
(see Ref \cite{feldman93}), suggests that the Berkeley survey may still be too
small to be seeing all of the large scale power.

     On large scales, the power spectrum is a better measure of clustering
power (i.e., has smaller errors) than the galaxy correlation function
\cite{peacock91}.  However, for scales $<100 h^{-1}$ Mpc, the power spectrum is
not as good an indicator.  Since the largest differences between the models
under consideration here occurs on smaller scales, we would like to supplement
our analysis with a different estimator of power on small scales.  To this end
we use the ``counts - in - cells" statistic from Efstathiou, et al., Ref.
\cite{efstathiou90}.  Here
the redshift space of IRAS galaxies is cut up into near cubical cells of
side length $\ell$ and the number of galaxies in
each cell is counted.  The cell to cell variance of galactic number is
a direct indication of the underlying density fluctuations of length $\sim
\ell$ in which the galaxies reside. The statistic is usually denoted by the
symbol $\sigma^2(\ell)$.  The values given in Ref.
\cite{efstathiou90} are
$\sigma^2(\ell)=$ $0.42\pm0.07$, $0.26\pm 0.05$, $0.21\pm 0.05$, and $0.047\pm
0.024$, for $\ell =$ 20, 30, 40, and 60
$h^{-1}$ Mpc, respectively.  They also give a value for $\ell = 10 h^{-1}$ Mpc,
but this represents a fluctuation which is strongly contaminated by non-linear
effects, and hence not appropriate for our analysis.  The QDOT counts in cells
data is presented in figure 3.

\subsubsection{POTENT bulk flow velocities}

  We also use the bulk flow velocities from the POTENT analysis \cite{dekel92}.
They represent the the rms velocities of spherical regions
of radius $R$.  The velocity has first been filtered with a Gaussian of width
$R=12 h^{-1}$ Mpc to reduce noise.  This data is shown in figure 4.  The
POTENT analysis has been done using the IRAS galaxies of the
Berkeley survey, and so suffers from the problem that enough of the universe
has
not been sampled to get a fair estimate of the velocities.  This is exacerbated
by the fact that velocities are even more sensitive to the very large scale
power than the mass fluctuation.  However, we would still like to find a way to
use these velocities because they do not depend on the bias.  (Velocities are
generated by the gravity of density fluctuations.)  Since we have velocities
from one local patch we will include the cosmic variance of the predicted
velocity for any given patch of the universe in our analysis.  Our treatment
will be described in the next section.

\subsection{The $\chi^2$ Test}

  The reduced $\chi^2$ statistics is calculated by comparing the set of $N$
predictions $y^{th}_i$ and observations $y^{obs}_i$ according to the following
formula
\begin{equation}
\chi^2 = {1\over N_{d.o.f.}}\sum_{i=1}^N \left({ y^{th}_i - y^{obs}\over
\sigma_i^{obs}}\right)^2,
\end{equation}
where $\sigma^{obs}_i$ are the standard observational errors.  $N_{d.o.f.}$
equals the number of observations $N$ minus the number of fitted theoretical
parameters in the model.  In our analysis we use the 36 values of the IRAS
$P(k)$, the 4 IRAS counts-in-cells, the 5 POTENT velocity values, the $b_I$
determination, and the COBE amplitude, which we will count as two points.  In
the $\chi^2$ analysis of $n=1$ models in
ref. \cite{taylor92}, the COBE data is counted as two points, one for the sky
noise at 10$^\circ$ and another for the value of $Q_{rms-PS}$.
Because of the complications of the non-gaussian beam pattern we have avoided
this procedure, but we still choose to weight the hexadecupole amplitude as if
it were two points in the analysis.  This forces the amplitude of density
perturbations to be slightly closer to the COBE normalized amplitude than by
weighting it as only one point.  The results of the analysis is quite similar
if we were to weight COBE as only one data point.  Thus we have a total of
$N=48$ data points which contribute to the $\chi^2$ sum.

   For each value of $n$ and $\Omega_{HDM}$ we find the values of $b_8$ and
$b_I$ which minimize $\chi^2(b_8, b_I)$ for each model.  This is effectively a
least squares fitting of $b_8$ and $b_I$ for each model, so the number of
degrees of freedom is 2 less than the number of points, $N_{d.o.f.} = 46$ in
our $\chi^2$ formula.  The confidence levels of rejecting the hypothesis that
the data
fits the model are found by integrating the normalized $\chi^2$ distribution.
The probability of getting $\chi^2>1.38$ is $<5$\%, so we can reject such
models with 95\% confidence.  We will now proceed to discuss how we compute
the $y_i^{th}$.

\subsection{Model Predictions From Linear Theory}

    Here we will explain how we compare our theoretical power spectra
$P_{th}(k)$ to the data presented in the previous section.  First of all, we
can calculate the hexadecupole $\delta T_4$ (measured by
COBE) with the formula given in Ref. \cite{anwcmbr} for the coefficient of the
fourth spherical harmonic for each of the power spectra we are considering.
However, we can find an effective wavenumber $k_{eff}$ for which the amplitude
of $P(k_{eff})$ is directly proportional to the amplitude of the hexadecupole.
We have found that the value $k_{eff} = 1.05 \times 10^{-3} h$/Mpc accurately
characterizes the hexadecupole moment for the limited range of $n$ we are
studying ($0.70<n<1.00$).

    The COBE experiment cannot distinguish between temperature fluctuations
generated by the Sachs-Wolfe effect of scalar (density) fluctuations and
tensor (gravity wave) fluctuations, both of which are products of inflation.
This signal confusion is worsened by the fact that the ratio of the
moments generated (at least on COBE scales) for scalar and tensor
contributions are nearly independent of scale
\cite{abbott86,crittenden93,turner93}.  The
overall amplitudes of the moments for scalar and tensor modes, however, can
be quite different, and one must construct a specific model of inflation
and then evaluate the density perturbation and gravity wave amplitudes
during inflation.  In some models of inflation, such as ``new" inflation
\cite{inflrev},  the contribution of gravity
waves to the COBE signal is negligible, and the COBE signal relates only to the
scalar density perturbations.

   Models of inflation which produce a significant amount of gravity waves
cannot be summarized with a universal formula.  However, there is a relation
derived from the toy model ``power law inflation" \cite{anwgw} which relates
the gravity wave (tensor) contribution to the  COBE signal $(\Delta
T/T)_T$ to the contribution from the density (scalar) fluctuations signal
$(\Delta T/T)_S$ via the spectral exponent of the density power spectrum $n$
\cite{liddle93,crittenden93,souradeep93}:
\begin{equation}
{(\Delta T/T)_T^2 \over (\Delta T/T)_T^2} \approx
7(1 - n).
\end{equation}
This relation also is reasonably accurate for chaotic inflation
models.  Since the COBE signal is a quadrature sum of the multipole moments,
decreasing
$n$ decreases the fraction of COBE signal due to density waves, and thus
implies a smaller amplitude for $P_{th}(k)$.   Thus the amplitude of the COBE
quadrupole moment is $\sqrt{8-7n}$ times larger than would be expected from
the density component alone.  We will consider models  with gravity waves
produced according to this formula.

   To simulate the IRAS power spectrum $P_I(k)$ we have to apply a couple of
correction factors to our theoretical $P_{th}(k)$.  First of all the
distribution of galaxies in redshift space  on large scales appears more
clustered because of the doppler contribution to the redshift from velocity
perturbations \cite{kaiser87}.  With a biased galaxy distribution ($b_I$) this
correction can be made by multiplying with the factor
\begin{equation}
 P(k) \rightarrow  \left[ 1 + {2\over 3 b_I} +{1\over 5 b_I^2}\right] P(k)
\end{equation}
However, on smaller scales there is the opposite effect; the doppler shifts
from the peculiar velocities of galaxies become large in comparison to the
Hubble velocities and in fact wash out this redshift clustering effect.  This
effect can be described with a velocity dependent factor \cite{peacock92}
\begin{equation}
 P(k) \rightarrow  P(k){\sqrt{\pi} \over 2}
{ {\rm erf}(k R_v)\over k R_v}
\end{equation}
where $R_v = 4.4\ h^{-1}$ Mpc for IRAS galaxies.

Applying these corrections to a linear power spectrum seems to give reasonable
agreement with the power spectrum of ``galaxies" in N-body simulations.  For
example, Feldman, et al. \cite{feldman93} present such a power spectrum (from
ref. \cite{klypin93}) for a model with 30\% HDM.  The N-body spectrum becomes
slightly higher than our linear spectrum for $k
{\mathrel{\raise.3ex\hbox{$>$\kern-.75em\lower1ex\hbox{$\sim$}}}} 0.15$
presumably due  to
non-linear corrections.  The worst disagreement
($ {\mathrel{\raise.3ex\hbox{$<$\kern-.75em\lower1ex\hbox{$\sim$}}}} 30$ \%),
occurs at the small wavelength end $k=0.2 h$ Mpc$^{-1}$ of the QDOT IRAS
$P(k)$.  This error is still much smaller than the QDOT error bars, so we
believe our procedure is relatively insensitive to non-linear effects.

   To compare our predictions with the counts in cells data we calculate the
mass variance in a spherical volume of radius $R= (3\ell^3/4\pi)^{1/3}$, i.e.,
the radius which encloses the same volume as $\ell^3$.  We have checked that
this procedure gives the same results as a cubic cell of length $\ell$ by
computing the variance in a cube and a sphere of equal volume using a CDM,
$n=1$ power spectrum.  The results were different by only a few percent.

   We calculate the counts in cells directly for each model, using the
same redshift space correction as appropriate for the IRAS power spectrum.
Peacock \cite{peacock91} has suggested a different method for comparing
counts-in-cells data to $P(k)$.  He noted that one can write
\begin{equation}
2 \pi^2 \sigma^2(l) = P(k_{eff})k_{eff}^3
\end{equation}
where the value of $k_{eff}$ depends on the spectrum.  This is the method used
by Taylor and Rowan-Robinson \cite{taylor92} for their $\chi^2$ analysis.
It is instructive to calculate $k_{eff}$ for a typical
model just to illustrate the wavenumbers being probed with the counts-in-cells
measurements.  For a $n=1$, $\Omega_\nu = 0.25$ model, the values for
$k_{eff}$ are given in Table 1.


Thus we see that for accurate count-in-cells numbers for these volumes, one
needs to accurately get the power spectrum out to
$k {\mathrel{\raise.3ex\hbox{$<$\kern-.75em\lower1ex\hbox{$\sim$}}}} 0.02$.

   Note we are not considering the datum from cells of length $\ell = 10
h^{-1}$
Mpc, as this point is strongly in the non-linear regime.  In appendix B we
estimate that even for $\ell = 20 h^{-1}$ Mpc, the non-linear effects may be
as large as $\sim$ 30 \%, which is roughly the same size as the disagreement
between our linear $P(k)$ and the N-body simulation $P(k)$ at the smallest
wavelength we consider, $k=0.2h/$Mpc.  For this reason we do not consider
estimates of the power on scales
much below $20 h^{-1}$ Mpc.  (The largest wavenumber from the IRAS QDOT $P(k)$
is $k_{max} = 0.195\ h$ Mpc$^{-1}$.)

   To do the testing, we first must calculate the linear theory power
spectra for all models with a primordial power spectral index $0.70 < n < 1.00$
and $0 <\Omega_\nu <0.5$.  We have limited $n$ to be $\leq 1$ because we are
considering only grand unified models of ``new" and ``chaotic" inflation.  We
calculate the power spectra for these models in steps of
$\delta n = 0.02$ and $\delta \Omega_\nu = 0.05$ using our form for $P(k)$:
\begin{equation}
P_{th}(k) = A k^n \left[ T(k) \right]^2,
\end{equation}
We will vary the amplitude $A$ by a factor of $2^{\pm 1}$ away from the COBE
implied central value in 20 logarithmically spaced steps.  However, since the
definition of $A$ is author dependent, we will discuss our results in
terms of the parameter $b_8$ for each model.  Thus we are effectively varying
$b_8$ in a range centered on the COBE best fit $b_8$.

The value of $b_I$ is somewhat more constrained than values of $b_8$.   For the
IRAS galaxies, separate determinations of $b_I$ have been made by comparing
their velocities and distributions.  These dynamical tests yield the 95\%
confidence values values $b_I = 1.16\pm 0.42$  ( Ref. \cite{kaiser91}) and $b_I
= 1.23\pm 0.46$ (ref. \cite{rr91}).  However,  the POTENT analysis
\cite{dekel93} of the Berkeley IRAS galaxies finds that  the 95\% confidence
interval for $b_I = 0.5 - 1.3$.   We therefore combine these measurements to
say that $b_I = 1.1\pm 0.3$ to cover the 95\% confidence overlap region of the
$b_I$ determinations.   We consider the measurement of $b_I = 1.1\pm 0.3$ to be
a  {\it bona fide} data point which we add to our $\chi^2$ test.   As is usual,
we will assume a constant bias factor independent of scale  and we will allow
$b_I$ to  vary in 7 linearly spaced steps between 0.8 and 1.4.  We calculate
the  $\chi^2$ for each of the values of $b_I$ and $b_8$, and find the the set
($b_I$, $b_8$) which gives the lowest value of $\chi^2$, and plot this  minimum
$\chi^2$ in our $\Omega_\nu-n$ contour plots.  Thus we allow each  model to put
its best face forward in our test.  Some of these models with  $b_I = 1.00$ are
shown in figure 5.

   We calculate the POTENT rms velocities by the same procedure as described in
ref. \cite{ss93}.  In our comparison of the rms POTENT velocities to
theoretical rms velocities, we will incorporate the cosmic variance of the
velocity field.   The idea is that with a Gaussian density field, the velocity
field will  also be Gaussian.  The magnitude of the velocity vector will have a
$\chi^2$ distribution with 3 degrees of freedom.  The variance of the rms
velocity is much larger than the POTENT velocity errors. For example the 68\%
theoretical confidence range on the average predicted velocity magnitude
$\langle v\rangle$ corresponds to ($\langle v\rangle - 0.48\langle v\rangle$,
$\langle v\rangle + 0.32 \langle v\rangle $), while the POTENT errors are
less than $\sim 15$ \%.  This is plotted in figure 4 for the $n=0.96$, 25\%
HDM model.  If we normalize an $n=1.00$ power spectrum to COBE then the
predicted velocities are within the POTENT 1 $\sigma$ error bars, regardless of
the HDM fraction (see e.g., $n=1.00$, 50\% HDM model in figure 4).  As we
decrease $n$ below 1.00, we will decrease the predicted velocities
well below the POTENT values.  Thus the upper limit on the predicted velocity
will be the most relevant quantity for comparing theory to observations.  We
combine this upper limit in quadrature with the POTENT 1 $\sigma$ error, and
use this as a fairer estimate of the error bar in our $\chi^2$ analysis.

While adding errors in quadrature is strictly correct only for Gaussian errors,
the velocity distribution is not too far from a Gaussian.  We plot several
theoretical predictions for $\langle v\rangle$ against the POTENT velocities
in figure 4.
Even with these huge error bars, we find that the predicted velocities for
models with significant gravity wave contributions will still have trouble
matching the POTENT derived velocities.

\subsection{ Results of Linear Data Analysis}

    The results of our $\chi^2$ test are presented in figures 6 and 7 for
models without and with gravity waves, respectively.  Starting with our
absolute best fitting model $n=1.00$, 30\% HDM and working outward, we plot
9 concentric curves with confidence levels of .5, 1, 5, 10, 25, 50, 68, and
95 \%.  (For reference, the best fit model has formally a probability
$4\times 10^{-5}$ for getting such a good fit by chance - although such small
probabilities are meaningless for statistical analysis.)
When one reaches the level of 1\%, areas where the theory does not match the
data become discernable in the data plots.  The first thing to note about the
graphs is that the overall level of $\chi^2$ for models with $n\sim 1.00$ is
quite low.  This result supports earlier claims that  $n=1$ models normalized
to COBE have sufficient power to explain ``large  scale power" apparent in
galactic clustering measures.  If the measurement  errors were smaller our test
would be a much better discriminator between  models.

We have two scenarios to discuss: models
with and without gravity waves.  First we point out one general trend which is
common to all models, and which does not show up in the $\chi^2$ numbers.  As
we decrease $n$ we decrease the amount of mass clustering power on small
scales.  To increase the galactic clustering power to compensate for  this, one
needs to increase the amount of galactic biasing $b_I$.  Since our program
finds the best fit $b_I$, we will point out that small $n$  models correspond
to highly biased models (large $b_I$).  We could better limit the models if we
had more precise information about $b_I$.

   The $\chi^2$ contours for models with no gravity waves are shown in  figure
6.  As can be seen we can rule out models with
$n {\mathrel{\raise.3ex\hbox{$<$\kern-.75em\lower1ex\hbox{$\sim$}}}} 0.7$ at
95\%
confidence.  There is a slight dependence of $n$ on the HDM fraction, with the
limit being $n\geq 0.67$ for models with no HDM and $n\geq 0.72$ for models
with 50\% HDM.  This is easily understood since models with a lot of ``tilt"
already have little galaxy scale power, and large HDM fractions exaggerate this
behavior.   The region of best fits occurs in a roughly rectangular  region
$0.85 {\mathrel{\raise.3ex\hbox{$<$\kern-.75em\lower1ex\hbox{$\sim$}}}} n
{\mathrel{\raise.3ex\hbox{$<$\kern-.75em\lower1ex\hbox{$\sim$}}}} 1.00$ and
$0.1 {\mathrel{\raise.3ex\hbox{$<$\kern-.75em\lower1ex\hbox{$\sim$}}}}
\Omega_{HDM} {\mathrel{\raise.3ex\hbox{$<$\kern-.75em\lower1ex\hbox{$\sim$}}}}
0.5$ That these
fits are quite good can be seen from the direct comparisons to data shown in
figures 3, 4, and 5 for a few models.

    The $\chi^2$ contours for models with gravity wave contributions according
to equation  2.5 are shown in  figure 7.  It is readily apparent that the
allowed parameter space is much smaller.  Here the 95\% confidence limit on $n$
rules out models with
$n {\mathrel{\raise.3ex\hbox{$<$\kern-.75em\lower1ex\hbox{$\sim$}}}} 0.85$.
Again, there is a slight dependence of
$n$ on the HDM  fraction, with the limit being $n\geq 0.83$ for models with no
HDM, and $n\geq 0.87$ for models with 50\% HDM.  The region of best fits occurs
in a roughly  rectangular region
$0.94 {\mathrel{\raise.3ex\hbox{$<$\kern-.75em\lower1ex\hbox{$\sim$}}}} n
{\mathrel{\raise.3ex\hbox{$<$\kern-.75em\lower1ex\hbox{$\sim$}}}} 1.00$,
$0.1 {\mathrel{\raise.3ex\hbox{$<$\kern-.75em\lower1ex\hbox{$\sim$}}}}
\Omega_{HDM} {\mathrel{\raise.3ex\hbox{$<$\kern-.75em\lower1ex\hbox{$\sim$}}}}
0.5$.  Including significant amounts of gravity waves
forces the normalization of the density power spectrum to be much lower, which
depletes the amount of clustering  power.

     We note that for our 95\% confidence limit an amount
${\mathrel{\raise.3ex\hbox{$>$\kern-.75em\lower1ex\hbox{$\sim$}}}} 50$ \% of
the
COBE signal is attributed to the effect of density fluctuations.  This
tells us that to fit large scale structure one requires that the COBE signal
cannot be dominated by inflation generated gravity waves.  Our best fit models
are those in which at least 80\% of the COBE signal is due to density
fluctuations.

  Our two conclusions from this analysis are that the power spectrum must be
close to the Harrison-Zeldovich form ($n=1.00$) and that the COBE
signal must be mostly due to density fluctuations.

\section{Constraints from Data on Non-Linear Structures}
\label{sec:level3}

    As noted in the previous section, we find the best models are those which
have $\Omega_{HDM}
{\mathrel{\raise.3ex\hbox{$>$\kern-.75em\lower1ex\hbox{$\sim$}}}} 0.1$.  The
statistical confidence of this conclusion is not very high, on the order
of 10\% for a given value of $n$.  This is not surprising,
since we have confined ourselves to only the largest scales,
$k {\mathrel{\raise.3ex\hbox{$<$\kern-.75em\lower1ex\hbox{$\sim$}}}} 0.2 h$
 Mpc$^{-1}$, where the transfer functions for all the models do not differ too
much (see figure 1).  The best place to discriminate between these models
is on smaller scales, where we have data only on the non-linear part of the
power spectrum.

\subsection{The Data}

    The relation between non-linear structures and the amplitude of the linear
power spectrum is quite complicated, and it is non-trivial to extract strong
conclusions from this data.   We will consider two such constraints which we
feel can be used relatively safely - constraints on the amplitude of $\sigma(8
h^{-1}$ Mpc) or equivalently $b_8$, and the requirement that quasars form early
enough to be compatible with observations.   The quasar constraint is a lower
bound on the power spectrum amplitude while  $\sigma(8 h^{-1}$ Mpc) is an upper
bound (at a somewhat larger scale).

\subsubsection{High redshift quasars}

   The discovery of quasars with high redshifts (about 20 with $z\geq 4$) was a
direct challenge to theories of structure formation.  One needs a minimum
amplitude for density fluctuation on galactic scales to account for the quasar
population.  Efstathiou and Rees (ref. \cite{er88}) considered the formation of
quasars in a highly biased ($b_8 = 2.5$) $n=1$ CDM model.  The basic
strategy is that in order for a massive black hole to form and power the quasar
emission, one first requires a  host dark matter halo to supply the
gravitational potential to induce baryonic infall.  The number density of
structures for a given power spectrum can be computed using the Press-Schecter
or BBKS techniques.  The number density of structures depends exponentially on
a parameter $\nu$ given by
\begin{equation}
\nu = {\sigma_c\over \sigma(R)}
\end{equation}
where $R$ is the radius of the initial collapsing region appropriate for the
objects in question and $\sigma_c$ is the linear theory value of
$\sigma(R)$ which corresponds to the gravitational collapse and virialization
of the object.  Using the spherical collapse approximation (see Appendix A)
$\sigma_c =3(12\pi)^{2/3}/20 = 1.69$.
In this application we use the Gaussian filtered mass fluctuation $\sigma_G(R)$
\begin{equation}
\sigma_G(R)^2 = {1\over 2 \pi^2} \int {dk} k^2 P(k) e^{-k^2 R^2}
\end{equation}
where the subscript ``G" is to remind us that we are using a Gaussian filtering
function.  The parameter $\nu$ then has the
physical meaning of the ratio of the overdensity in a collapsed object to the
ambient rms overdensities.  Thus specifying $\nu$ tells us the amplitude of the
density fluctuations on the scale $R$.  Also, since the calculated number
density depends exponentially on $\nu$, it is hoped that errors in the measured
number density will not significantly change the value of $\nu$.

   Efstathiou and Rees (ref. \cite{er88}) estimated that the minimum mass
for a quasar halo was $\geq 2\times 10^{12} M_\odot$.  They found that a $b_8 =
2.5$, $n=1$ CDM model could account for the number density of quasars at
least out to a redshift of $z=5$.  Haehnelt and Rees (ref. \cite{haehneltrees})
improved  on this treatment by showing that the $b=2.5$ CDM could fit the
quasar luminosity function at a variety of redshifts.  Since both
increasing the HDM fraction and reducing $n$ lessen the amount of power on
small scales, it is
obvious to ask whether the amplitude of quasar scale fluctuations has
decreased below that required for making quasars.  Haehneldt (ref.
\cite{haehnelt93}) used the  techniques of ref. \cite{haehneltrees} to limit
the
HDM fraction in $n=1$ models and $n$ in CDM models, finding that $n>0.75$ in
CDM models and the HDM fraction must be $\Omega_{HDM}\leq .3$.   Schaefer and
Shafi (ref. \cite{ss93}) showed that 25\% HDM and $n=0.94-0.97$ is compatible
with the quasar density out to a redshift of  $z\sim 5$.

   The Press-Schecter technique is, however, not without errors.  Even if we
knew the quasar number density perfectly, and hence could deduce the exact
value  of $\nu$, there would still be uncertainties.  These uncertainties have
been discussed before in a variety of places.  Here we follow the discussion of
reference \cite{liddle93}.  The uncertainties in $\nu$  can be found by
\begin{equation}
\Delta {\rm ln} \nu = \Delta {\rm ln} \sigma_c - {d\sigma(R)\over dR}
{dR\over dM} \Delta {\rm ln}M
\end{equation}
First of all it is not clear what value of $\sigma_c$ to use.  While most
authors use the value 1.69, comparison of Press-Schecter results to N-body
simulations imply values of $\sigma_c$ anywhere from 1.33 \cite{er88} to 1.69
\cite{white93}.  To bracket these values we can assume  $\sigma_c = 1.5\pm
0.2$.  Second there are errors in the quoted mass of the object.  The
theoretical value for the quasar halo mass is somewhat uncertain, and the mass
of the object derived from the luminosity function comes with an additional
error.  To evaluate the total uncertainty one needs to know the value of
$(d\sigma(R)/dR)(dR/dM)$.  For the quasar halo mass of $2\times 10^{12}
M_\odot$, the Gaussian filter radius is  $R=0.6 h^{-1}$ Mpc.  For $n=1$ models
$(d\sigma(R)/dR)(dR/dM)$ ranges from 0.14 for CDM models to 0.033 for 50\% HDM.
Assuming a factor of 3 error in the observational and theoretical masses, we
then have
\begin{equation}
\Delta {\rm ln} \nu = \pm 0.13 \pm 0.04 \pm 0.04
\end{equation}
for 50\%HDM ($n=1$) which has the smallest uncertainties of the $n=1$ models.
 We conclude that there could be an uncertainty of about
20\% in the amplitude of the density fluctuation $\sigma(R)$ derived from the
quasar density.

   In order to have the proper quasar density at early redshifts, one needs to
have $\sigma(0.6 h^{-1}$ Mpc$) \geq 1.1$ today (using $\sigma_c = 1.5$).   The
effect of the hot dark matter on the growth of density fluctuations at these
scales changes the above threshold value by only a few percent
\cite{haehnelt93}.  To forbid models which are unlikely to have quasars form
early enough, we make the replacement
\begin{equation}
\chi^2 \rightarrow \chi^2 + 20 \left[{\sigma(0.6 h^{-1}{\rm Mpc}) - 1.1 \over
0.22}\right]^2
\end{equation}
whenever $\sigma(0.6 h^{-1}{\rm Mpc}) \leq 1.1$.  We have weighted this
contribution to $\chi^2$ by 20 to strongly penalize models without sufficient
small scale power.  Since there are about 40 data points in our $\chi^2$
analysis an amplitude $\sigma(0.6 h^{-1}{\rm Mpc})$  which is 40\% ($2\times
20$\%) smaller than our threshold amplitude will  cause $\chi^2$ to be so large
that the model will be ruled out by the quasar data alone.

\subsubsection{$\sigma(8 h^{-1}$ Mpc) }

   Since it has become traditional to specify the amplitude of linear theory
by $b_8$, some attention has been paid to determining the value of $b_8$
from observations.  There are several ways of getting at this quantity which
seem to be converging on the range $b_8 = 1.5-2.0$ for the $\Omega=1$ models.
We will consider some of the more recent attempts to constrain this parameter.

    The easiest way to estimate the density fluctuations is by finding the
variance of the galactic number density.  For the IRAS galaxies we have two
estimates of the number density fluctuations: one from the QDOT survey
\cite{saunders92}
\begin{equation}
b_I \sigma^{non-linear}(8 h^{-1}{\rm Mpc}) = 0.69 \pm 0.09
\end{equation}
which is in perfect agreement with the estimate from the Berkeley survey
\cite{fisher93}
\begin{equation}
b_I \sigma^{non-linear} (8 h^{-1}{\rm Mpc}) = 0.69 \pm 0.04
\end{equation}
The error on the averaged combined measurement is essentially the same as the
Berkeley error, $b_I \sigma^{non-linear} (8 h^{-1}{\rm Mpc}) = 0.69 \pm 0.04$.
The 95\% confidence upper limit is then $b_I \sigma^{non-linear}
(8 h^{-1}{\rm Mpc}) \leq  0.77$
If we take an extremely small value for $b_I = 0.5$ which is the POTENT 95\%
confidence lower limit, this would imply that
$ \sigma^{non-linear} (8 h^{-1}{\rm Mpc}) \leq 1.54$.   If we use the spherical
collapse model (see Appendix B) to determine what that means in terms of the
linear density fluctuation, this implies
\begin{equation}
\sigma^{linear} (8 h^{-1}{\rm Mpc}) \leq 0.71,\ {\rm or}\ b_8 \geq 1.40.
\end{equation}
To be extra cautious, we will adopt the constraint
\begin{equation}
\sigma^{linear} (8 h^{-1}{\rm Mpc}) \leq 0.80,\ {\rm or}\ b_8 \geq 1.25
\end{equation}
for a new round of $\chi^2$ testing.

   A few remarks are in order here concerning the value of $b_8$.
Early investigations found that CDM models with  $n=1$ required values of $b_8
\sim 2-3$ \cite{defw85} to get agreement with  galactic velocity dispersion
data and correlation functions.  However,  large scale structure data required
that CDM have smaller values of $b_8$.   It was postulated that dynamic
effects, especially ``velocity bias"  \cite{couchman92} might accomodate
smaller values of $b_8$.   High resolution simulations \cite{gelb93} however,
find $b_8\geq 1.4 $, despite their confirmation of the existence of the
``velocity bias" effect.  This conclusion seems to also  hold for $n<1.00$
\cite{gelb92}.  Ref. \cite{gelb92} also suggests that this is  true even in
models with C+HDM.  However, they did not consider the effect of the dynamical
effects of HDM in their study, and this seems to allow us to use $b_8<2$ with
significant amounts of HDM \cite{davis92,klypin93,jing93}.   No systematic
study has been done to determine what values of $b_8$ are allowed, although
$b_8=1.5$ seems to work with $n=1.00$, $\Omega_{HDM} = 0.30$.  Our constraint
$b_8 \geq 1.25$ seems to be easily consistent with these case studies.

   Another constraint on the mass fluctuation amplitude comes from cluster
properties.  Since clusters are a few Mpc in size, they are an almost ideal
choice for determining $b_8$.  However, to determine the number density of
cluster mass structures analytically, one must use a procedure such as the
``BBKS" method \cite{bbks} or the Press-Schecter \cite{ps} model.   There are
some uncertainties associated with this technique however, which we have
discussed in the previous section on quasars.  In ref. \cite{vdns92} it was
found that $b_8 = 1.12-0.96$  for models with 0\% - 50\% HDM,  based on $R\geq
0$ Abell cluster number abundance.  To arrive at this number van Dalen and
Schaefer (ref. \cite{vdns92}) used the spherical collapse model.  The $R=0$
Abell  clusters are the poorest Abell clusters and represent initially weaker
density  fluctuations.  The smaller amplitude density fluctuations tend to be
highly asymmetric \cite{bbks} and collapse faster than spherical perturbations
(see  {\it e.g.} ref. \cite{bertschinger93}), so a  spherical collapse model
probably gives $b_8$ values which are too small.  It was  estimated that more
accurate values for $b_8$ would be at least 30\% larger,  i.e., $b_8 = 1.46 -
1.25$, consistent with the adopted restriction $b_8
{\mathrel{\raise.3ex\hbox{$>$\kern-.75em\lower1ex\hbox{$\sim$}}}} 1.25$.

   Perhaps a more reliable way to study the mass fluctuations in clusters is to
select them by their X-ray temperatures, as this gives a direct indication of
the gravitational mass potential.  These studies tend to give  results
consistent with much higher $b_8$, implying $b_8\sim 2.0-2.5$  (\cite{frenk90},
$b_8=1.6-1.9$ \cite{white93} and $b_8=2.0$, \cite{bartlett93}.  Thus we find
our restriction $b_8\geq 1.25$ is, if anything, too conservative.

\subsection{Results of Non-Linear Analysis}

In Figures 8 and 9 we again plot the $\chi^2$ contours for models with and
without significant gravity wave temperature anisotropies.  The effect of our
non-linear constraints is clear.  The restriction $b_8\geq 1.25$ forces the
normalization of small $\Omega_{HDM}$, $n\sim1$ models to be too low to
match the large scale structure data.  This is symptomatic of the CDM models
which, when normalized to COBE, have too much small scale power.  The effect of
enforcing the lower limit on the amplitude from quasars is to cut out a
triangle of parameter space corresponding to large $\Omega_{HDM}$ and small
$n$.   This is symptomatic of $\Omega_{HDM}\sim 1$ model problems, there being
not enough small scale power to explain the early epoch of quasar formation.

   What we are left with is a patch of parameter space which has $n\sim 1$ and
$\Omega_{HDM} = 0.3 \pm 0.2$.  This is in agreement with earlier studies and
shows that models with $\Omega_{HDM}\sim 1/4$ are significantly better fits
to the data than with no HDM.  The range of $n$ is roughly the same as in the
linear analysis, although the non-linear quasar constraint has effectively
chopped off the low $n$, high HDM fraction corner of parameter space of the
previous best fits.  Our allowed region of parameter space overlaps the allowed
region found by Liddle and Lyth, (ref. \cite{lyth93}).  However, their
analysis took the non-linear constraints of refs. \cite{haehnelt93,white93}
at face value so their allowed region is somewhat smaller than ours.
They noted however, that their allowed region was meant to be suggestive of
trends in the data and should not be taken literally.  On the other hand we are
taking pains to be overconservative with non-linear constraints in the hope
that our limits will be firmer.

   Thus we find the following properties which it is desirable for inflation to
have.  We require a density perturbation spectrum which is quite close to a
Harrison-Zeldovich spectrum as the data do not seem to favor much ``tilting".
For models with a negligible gravity wave contribution to the COBE signal,
the best fits occur for
$0.9 {\mathrel{\raise.3ex\hbox{$<$\kern-.75em\lower1ex\hbox{$\sim$}}}} n
{\mathrel{\raise.3ex\hbox{$<$\kern-.75em\lower1ex\hbox{$\sim$}}}} 1.0$.  In
models with some gravity
wave contribution, we find an even tighter range of best fit $n$ values, namely
$0.94 {\mathrel{\raise.3ex\hbox{$<$\kern-.75em\lower1ex\hbox{$\sim$}}}} n
{\mathrel{\raise.3ex\hbox{$<$\kern-.75em\lower1ex\hbox{$\sim$}}}} 1.00$.
Since we used equation 2.5 to determine this, we
see that for $n=0.94$ the gravity wave contribution to COBE is only 20\%.  Thus
the data favor models for which density perturbations are
${\mathrel{\raise.3ex\hbox{$>$\kern-.75em\lower1ex\hbox{$\sim$}}}} 80$\%
responsible for the COBE signal.  We also find
$0.15 {\mathrel{\raise.3ex\hbox{$<$\kern-.75em\lower1ex\hbox{$\sim$}}}}
\Omega_{HDM} {\mathrel{\raise.3ex\hbox{$<$\kern-.75em\lower1ex\hbox{$\sim$}}}}
0.45$ gives the closest fits to the data,
which implies we would nominally like one flavor of neutrino with a
mass $m_\nu = 3 -10$ eV. (The best fits imply a narrower range $\Omega_{HDM} =
0.20-0.35$.   With these attributes in mind we proceed to explore
possible models for inflation.

\section{Models of Inflation}

\def\strut{\rule[-.5cm]{0cm}{1cm}}

    Grand unified theories (GUTS) provide the simplest framework for
implementing the inflationary scenario. Although supersymmetric GUTS are
currently more popular, for completeness we will also consider the ordinary
non-supersymmetric versions. The simplest example of the latter is provided by
$SO(10)$ with an intermediate mass scale. The minimal non-supersymmetric
$SU(5)$ model is excluded both by the precise determination of $\sin^2\theta_W$
and by proton decay experiments. This is just as well from our point of view
since, as observed a decade ago, \cite{holman83}, non-SUSY $SO(10)$ models with
an intermediate (B-L breaking) scale $M_{B-L} (\sim 10^{12}\ GeV)$ strongly
suggest that the tau neutrino mass is in the $eV$ range. The presence of a
$U(1)$ axion symmetry not only resolves the strong CP problem but also
provides the cold dark matter component.

Two versions of the inflationary scenario, `new' and `chaotic',  are readily
realized in GUTS. The spectral index $n$ of density  fluctuations in the
simplest realistic models typically lies between 0.96 and 0.92, although in
some versions of  chaotic inflation with SUSY GUTS, $n$ could be as low as
0.88. Values of $n$ much smaller than this are not particularly well motivated,
both from the point of view of model building as well as observations of the
large scale structure.

\subsection{Inflation with Non-Supersymmetric SO(10)}

For definiteness, let us consider the following breaking:

$$\begin{array}{ccccc}
SO(10) & \longrightarrow & SU(4)_c \times SU(2)_L \times SU(2)_R &
\longrightarrow & SU(3)_c \times SU(2)_L \times U(1)_Y\\
& M_X & & M_{B-L} &\end{array}$$

\noindent
A recent two loop renormalization group calculation involving the
gauge couplings gives \cite{deshpande92} $M_X \sim 10^{15} - 10^{16}\ GeV$
and $M_{B-L} \sim 10^{12\pm 1}\ GeV$, consistent with the measured values of
$\alpha_c (M_Z)$ and $\sin^2\theta_W (M_Z)$.

A simple version of the see saw mechanism for neutrino masses
\cite{gellman79} suggests the hierarchy:

\begin{equation}
m_{\nu_1} : m_{\nu_2} : m_{\nu_3} \approx m^2_u : m^2_c: {\cal
O}(10^{-1}) m^2_t
\end{equation}

\noindent
Here the three mass eigenstates $\nu_1, \nu_2, \nu_3$ primarily
consist of $\nu_e, \nu_\mu$ and $\nu_\tau$ respectively, and we have
assumed (see later) that the heavy (right handed Majorana) neutrino associated
with the third family is a factor 10 (or so) heavier than the other
two.

The MSW interpretation of the solar neutrino data suggests \cite{mswn92}
that the $\nu_2$ mass is $\sim 10^{-2.7} - 10^{-2.5}\ eV$. With $m_t (m_t)
\sim 130 - 150\ GeV$, as suggested by recent analyses of the
electroweak data, we expect the $\nu_3$ (essentially $\nu_\tau$, with
some admixture of $\nu_\mu$ and $\nu_e$) mass to be $\sim 5 - 10\
eV$. (Without the numerical factor in equation (4.1), this mass would exceed
the cosmological bound.)  Note that according to this simple $SO(10)$
example, unless the
$\nu_\tau - \nu_\mu$ mixing happens to be tiny, the two neutrino
oscillation experiments CHORUS and NOMAD should determine whether or not the
`tau' neutrino is cosmologically significant.

  The $SO(10)$ model with the above symmetry breaking chain suggests the
existence of some dark matter in the form of tau neutrino. However, two
essential ingredients are still missing, implementation of inflation and a
candidate for cold dark matter (CDM). [Recall that inflation with only hot dark
matter (HDM) does not seem compatible with the observed large scale structure,
especially galaxy formation.] The simplest way to incorporate CDM here is to
invoke a $U(1)$ axion symmetry  \cite{peccei77} broken at a scale around
$10^{11} - 10^{12}\ GeV$. Both the axions and neutrinos are then cosmologically
significant. It may be useful to reiterate how this has come about. In
non-supersymmetric $SO(10)$, an intermediate scale is needed to bring about
consistency with the measured value of $\sin^2\theta (M_Z)$ as well as with the
lower limits on the proton lifetime. This forces the gauged $B-L$ symmetry to
break at an intermediate scale which, for the above chain,  is about $10^{12\pm
1}\ GeV$. Coupled with the see saw, this strongly suggests that the tau
neutrino mass $(\sim m^2_t/M_{B-L})$ is in the $eV$ range. That is, the
neutrino is a significant component of the dark matter. Cold dark matter is
then needed to reconcile the inflationary scenario with observations related to
large scale structure.

As far as inflation is concerned, the most straightforward scenario is
realized by introducing a weakly coupled gauge singlet field $\phi$ a la
Shafi and Vilenkin \cite{shafi84}. The part of the potential which drives (new)
inflation is given by

\begin{equation}
V(\phi) = \lambda \phi^4 \ln ( \frac{\phi^2}{M^2} - \frac{1}{2})
\end{equation}

\noindent
where $M$ denotes the vacuum expectation value (vev) of $\phi$. The quantity
$\lambda$ can be reliably estimated by considering the contribution of the
scalar perturbations to the microwave background quadrupole anisotropy and
identifying it with COBE's determination of $Q_{rms - PS}$. [Note that for the
potential in (4.2), the spectral index $n \approx 0.94$, and the tensor
contribution to the anisotropy is negligible.]

One has (the subscript $S$ denote the scalar contribution)

\begin{equation}
\left( \frac{\Delta T}{T} \right)^2_S \simeq \frac{32\pi}{45}
\frac{V^3}{V^{\prime^2} M_P^6} |_{k \sim H}
\end{equation}

\noindent
where $M_P (= 1.2 \times 10^{19}\ GeV)$ denotes the Planck scale, and the right
hand side is to be evaluated when the scale $k^{-1}$, corresponding to the
present horizon size, crossed outside the horizon during inflation. Equation
(4.3) can be rewritten as

\begin{equation}
|\Delta T/T|_S \simeq 0.067 \sqrt{\lambda} N_H^{3/2} |\ln
(\phi_H^2/M^2)|^\frac{1}{2}
\end{equation}

\noindent
where $N_H (\approx 55)$ denotes the number of $e$ foldings
experienced by this scale, and $\phi_H$ is the field value when the
scale crossed outside the horizon.

The logarithmic factor in (4.4) is of order 1 - 10, assuming that the
vacuum energy density that drives inflation is comparable to $M_X^4$.
For $M_X \sim 10^{15.5}\ GeV$, $M \sim M_P = 1.2 \times 10^{19}\
GeV$, and taking $(\Delta T/T)_{\rm COBE} \approx 6 \times 10^{-6}$,
the fundamental quantity $\lambda$ is estimated to be

\begin{equation}
\lambda \approx 2.2 \times 10^{-14}
\end{equation}

Any inflationary scenario is incomplete without an explanation of the origin of
the observed baryon asymmetry in the universe. In the present case the inflaton
mass $m_\phi \approx 10^{12.5}\ GeV$, and so the basic idea is to create an
initial lepton asymmetry via inflaton decay into one or more species of the
heavy (`right handed') majorana neutrinos. The appearance of `sphaleron'
induced processes at the electroweak scale converts a specified fraction of
this asymmetry into the observed baryon asymmetry. Details of how a
satisfactory scenario is realized along these lines can be found in ref.
\cite{lazarides91}.  We mention here a few salient features:

\begin{quote}
(i)$\;\;$ The reheat temperature $T_r \sim 10^{8.5}\ GeV$ so that the out
of equilibrium condition on the `heavy' ($\sim 10^{12}\ GeV)$
neutrinos is readily satisfied;

(ii)$\;\;$The requirement that for temperature below $T_r$ the rates
for lepton number violating
processes $(\nu\nu \leftrightarrow H^{\circ *} H^{\circ *}$ and $\nu
H^\circ \leftrightarrow \bar{\nu} H^{\circ *}$, where $H^\circ$ is
the electroweak scalar higgs) be smaller than the expansion rate $(H \simeq 20
T^2/M_P$, where
$H$ denotes the Hubble constant) of the Universe, leads to the
following constraint on the light neutrino masses \cite{harvey90}:
\end{quote}

\begin{equation}
m_\nu \stackrel{_<}{_\sim}  {4\ {\rm eV}\over
(T_r/10^{10}\ {\rm GeV})^\frac{1}{2} }\approx 20\ eV
\end{equation}

\begin{quote}
This is fully consistent with a cold plus hot dark matter scenario where
neutrinos in the mass range $3 - 10\ eV$ are needed.

(iii)$\;\;$In the present approach the colored scalar triplets
which mediate proton
decay are not needed for baryogenesis and consequently are allowed to
have masses $\sim M_X$.

(iv)$\;\;$Finally, it is possible, following ref. \cite{pi84}, to identify the
inflaton with the field that spontaneously breaks the axion
symmetry. This would make for a more economical approach.

(v)$\;\;$With an appropriate re-interpretation, the chaotic
inflationary scenario can be realized within the framework of this
$SO(10)$ model.
The ratio of the scalar to the tensor contribution in this case is

\begin{equation}
(\Delta T/T)^2_T /(\Delta T/T)^2_S \simeq 0.22.
\end{equation}

\end{quote}

\subsection{Supersymmetric Inflation}

The presently measured gauge couplings of the standard model, when
extrapolated to higher energies with supersymmetry (SUSY) broken at
scales around $10^3\ GeV$ \cite{amaldi91}, appear to merge at scales
around $10^{16}\
GeV$. This is a boost for supersymmetric GUTS, with $SU(5)$ or $SO(10)$
being the obvious gauge groups. In the presence of unbroken matter
parity, either of them can provide a cold dark matter candidate in
the form of LSP (lightest supersymmetric particle). However, hot dark
matter in the form of massive neutrinos most naturally appear in the
$SO(10)$ model. The supersymmetric $SO(10)$ scheme has some
additional features which make it attractive from the particle
physics viewpoint. For instance,

\begin{quote}
(a)$\;\;$A $Z_2$ subgroup of the center of $SO(10)$ (more
precisely Spin (10)) is left unbroken if tensor representations are
employed to do the symmetry breaking. This $Z_2$ symmetry \cite{kibble82},
which is not contained in $SU(3)_c \times SU(2)_L \times U(1)_Y$, acts
precisely as matter parity!

(b)$\;\;$In some versions of SUSY $SO(10)$, the important parameter
$\tan\beta (\equiv \phi^u/\phi^d$, the ratio of the two vevs which provide
masses to `up' type and `down' type quarks) is predicted to lie close to
$m_t/m_b$ \cite{ananth91}. One consequence of this is the identification of
the `bino' (the supersymmetric partner of the $U(1)_Y$ gauge boson) as the
LSP, with mass $\sim 200-300$ GeV.

(c)$\;\;$Fermion Mass Ansatzes have recently attracted a fair
amount of attention and are most simply realized within the framework
of SO(10) \cite{lazarides91b}.
\end{quote}

To summarize, particle physics considerations as well as observations
of large scale structure which favor a cold plus hot dark matter
scenario, together suggest SUSY $SO(10)$ as an attractive way to proceed.
Inflation, either `new' or `chaotic', can be implemented by introducing
a suitable singlet superfield.
Remarkably enough, singlets are typically employed to
achieve the breaking of the GUT symmetry, (without breaking SUSY) and we
exploit one of them to induce inflation!

Let $\Phi$ denote the $SO(10)$ singlet (inflaton) superfield,
$\chi(\bar{\chi})$ are the higgs superfields in the $126(\overline{126})$
representations whose vevs provide Majorana masses to the right
handed neutrino, and $16_i (i = 1,2,3)$ are the matter superfields. To
simplify the discussion, we restrict attention to the sector
involving an interplay only between these superfields. This allows us to
discuss the salient features of the (chaotic) inflationary scenario
including baryogenesis. Consider the renormalizable
superpotential

\begin{equation}\begin{array}{rl}
W = \alpha \Phi (\chi \bar{\chi} - M_X^2) & + \frac{M}{2} \Phi^2 +
\frac{\beta}{3} \Phi^3\strut\\
& + \gamma_{ij} 16_i 16_j \bar{\chi}\end{array}
\end{equation}

\noindent
Note that $\Phi \rightarrow \Phi$ under the matter parity contained
in $SO(10)$ (similarly $\chi, \bar{\chi} \rightarrow \chi,
\bar{\chi}$, and $16_i \rightarrow - 16_i$).

The superpotential $W$ gives rise to a supersymmetric ground state
in which (vevs refer to the scalar components of the superfields)

\begin{equation}\begin{array}{ccl}
<\phi> & = & 0,\; <16_i> = 0,\strut\\
<\chi> & = & <\bar{\chi}>^* \neq 0\end{array}
\end{equation}

\noindent
It is clear that matter parity is unbroken and we expect the LSP to
contribute to the cold dark matter component.

Even though $B-L$ is now broken at $M_X \sim 10^{16}\ GeV$, the right handed
`tau' neutrino mass must be of order $10^{12} - 10^{13}\ GeV$, if the `light'
tau is to be the dark matter component.  The inflaton must be at least twice
as heavy, and one simple way to have $m_\phi \sim 10^{13}\ GeV$ is to arrange
the coefficients $\alpha$ and $M$ in (4.8) to be of order $m_\phi/M_X$ and
$m_\phi$ respectively.  The decay rate of the inflaton into right handed
neutrinos is given by

\begin{equation}
\Gamma \sim \frac{1}{4\pi} \left( \frac{m_\phi}{M_X} \right)^6
m_\phi\ GeV
\end{equation}

\noindent
With $m_\phi \sim 10^{13}\ GeV$, the reheat temperature
$T_r$ is of
order $10^6\ GeV$. Baryogenesis via leptogenesis now proceeds along
the lines given in ref. \cite{lazarides91}.  Note that because of the
relatively low $T_r$, the otherwise vexing gravitino problem is neatly
avoided in this approach.

Depending on the details, the spectral index $n$ lies between 0.94 (if
the quartic potential dominates) and 0.96 (with a quadratic potential
dominant during the chaotic inflationary phase). The ratio
$(\Delta T/T)^2_T /(\Delta T/T)^2_S \simeq 0.22(0.11)$, respectively.

\subsection{Inflation Without the Singlet}

The question we wish to address here is the following: Is it possible
to implement inflation with GUTS without the gauge singlet? Surprisingly
perhaps \cite{lazarides93}, an affirmative answer appears possible for a
special class of supersymmetric GUTS in which, up to a normalization
constant, the GUT scale is determined in terms of $M_S$ and $M_P$, the
SUSY breaking scale and the Planck scale respectively. Moreover, the
normalization constant is fixed from the quadrupole anisotropy.
Such models \cite{candelas85} naturally arise after compactification of the ten
dimensional $E_8 \times E_8$ heterotic string theory \cite{gross85}, and models
based on $G \equiv SU(3)_c \times SU(3)_L \times SU(3)_R$ or its subgroups
provide some elegant examples. The scalar fields needed to spontaneously
break $G$ to $SU(3)_c \times SU(2)_L \times U(1)$ can be used to drive
inflation!

The key ideas are relatively straightforward and perhaps best illustrated by a
simplified example. Consider a rank five gauge symmetry $H \equiv
SU(3)_c \times SU(2)_L \times U(1)_Y \times U(1)^\prime$, where the
extra factor $U(1)^\prime$ is to break at some superheavy scale $M$
(below $M_P$).  Let $\phi, \bar{\phi}$ denote the pair
of higgs scalars whose vevs do this breaking. Note that
$<\phi>\ =\linebreak <\bar{\phi}>^*$ so that the $D$ term vanishes. Since the
only independent dimensionful parameters are $M_S$ and $M_P$ $(M$ is
determined in terms of them), the $\Phi \bar{\Phi}$ terms in
the superpotential are either absent or carry coefficients of order
$M_S$. (Here $\Phi, \bar{\Phi}$ denote the corresponding
superfields.) Moreover, in order to ensure $F$ flatness, the cubic
couplings $\Phi^3, \bar{\Phi}^3$ are also absent (otherwise
$U(1)^\prime$ would break at scales $\sim M_S$).

In the absence of SUSY breaking the superpotential $W$ is taken to be

\begin{equation}
W = h \chi \chi \Phi + \frac{\kappa}{M_P} (\Phi \bar{\Phi})^2 +
\cdots
\end{equation}

\noindent
where $\chi$ denotes some matter superfield with the coupling $h$ of
order unity. Assuming a radiative breaking scenario along the
lines envisaged in supergravity models with electroweak breaking,
the effective potential takes the generic form

\begin{equation}
V (\phi, \bar{\phi}) \sim - M^2_S |\phi|^2 + \kappa^2
\frac{|\phi|^6}{M_P^2}
\end{equation}

\noindent
Minimization of (4.12) leads to the result

\begin{equation}
M \equiv |<\phi>| = |<\bar{\phi}>| \sim \kappa^{-\frac{1}{2}}
(M_SM_P)^{\frac{1}{2}}
\end{equation}

Provided that $\kappa^2$ is sufficiently small, the potential in (4.12)
will yield a satisfactory (chaotic) inflationary scenario. It turns
out that $\kappa \sim 10^{-7}$ (from COBE), which gives $M \simeq
10^{15}\ GeV$. The spectral index $n \approx 0.92$, while the ratio
of the tensor to the scalar quadrupole anisotropy is $(\Delta
T/T)^2_T/(\Delta T/T)^2_S \approx 0.4$. Values of $n \approx 0.88$
(but no lower!) can be entertained within this framework. This is
just as well since our analysis in the earlier sections seems to
favor the range $1.0 \stackrel{_>}{_\sim} n \stackrel{_>}{_\sim}
0.9$.

To summarize, grand unification provides an elegant framework for
implementing both the `new' and `chaotic' inflationary scenarios.
Some popular models based on $SO(10)$ or $SU(3)_c
\times SU(3)_L \times SU(3)_R$ predict the value of the spectral
index $n$ in the range 0.96 to 0.92. Cold dark matter, in axions
and/or LSP, as well as hot dark matter in massive neutrinos are
readily incorporated in these schemes.

\section{Conclusions}

  We have performed a $\chi^2$ goodness of fit test of the predicted linear
theory power spectra against data on scales ranging from 1 to $10^4$ Mpc,
mainly from the COBE satellite and the QDOT IRAS survey of galaxies.  We
find that the inflation based scenario of large scale structure formation, in
which the dark matter consists of cold plus hot components, can provide a
good fit to large scale structure data.

   Taking the primordial power spectrum to have spectral exponent $n$, we find
with 95\% confidence, respectively, that
$n {\mathrel{\raise.3ex\hbox{$>$\kern-.75em\lower1ex\hbox{$\sim$}}}} 0.7$,
($n {\mathrel{\raise.3ex\hbox{$>$\kern-.75em\lower1ex\hbox{$\sim$}}}} 0.85$),
in models with (without) significant gravity wave contributions to the COBE
anisotropy.  The precise bound depends slightly on the HDM fraction.  We
find in models with a significant gravity wave anisotropy, that the COBE signal
must not be dominated by the gravity wave contribution.

  If one insists on only one type of dark matter, {\it i. e.}, CDM, the best
fits are for $n=0.84$ ($n=0.92$) in models without (with) gravity waves,
respectively.

   The best fit region for all data, including some constraints from non-linear
structure, is an roughly an ellipse (see figures 8 and 9).  For models with
small amplitude gravity wave anisotropies, the focii of the ellipse are
approximately at ($\Omega_{HDM} = 0.20$, $n=0.92$) and ($\Omega_{HDM} = 0.35$,
$n=0.98$).  For models with a large tensor COBE anisotropy, the ellipse is
more eccentric, with the focii at roughly ($\Omega_{HDM} = 0.2$, $n=0.96$)
and ($\Omega_{HDM} = 0.35$, $n=0.98$).  Thus, the best fits occur for
$\Omega_{HDM} \sim 0.20-0.35$, with $n$ very close to unity.

Realistic  examples of inflation from grand unification theories,
including both supersymmetric and ordinary GUTS, which have these properties
have been presented.  These models are also consistent with other cosmological
and particle physics constraints.

\acknowledgments

We wish to thank Hume Feldman for supplying us with
the IRAS $P(k)$ and Michael Strauss for interesting discussions.  This work is
supported by grants from NASA (NAGW - 1644) and DOE (DE-FG02-91ER40626).

\appendix
\section{Transfer Function Calculations}

As described in ref. \cite{sss}, we integrate the
Fourier space evolution equations in
(conformal) time using the gauge-invariant variables
for the density $\Delta_{ca}$ and velocity $V_a$ perturbations in each
energy density component (CDM, neutrinos, photons, baryons) as given in ref.
\cite{ks84}.  Here we integrate the equations in conformal time
instead of the scale factor as we had done previously. We begin the integration
well before the matter dominated era ($z = 8.3\times 10^6$ )
and integrate up until the present time ($z=0$).
We have used only one flavor of massive neutrino, with the other two flavors
essentially massless.  For the baryon equations we use the equations for the
difference between the baryon and photon density and velocity perturbations
(the variables $S_{br}\equiv \Delta_{c,baryon}- (3/4)\Delta_{cr}$ and
$V_{br}\equiv V_b - V_r$
given in ref. \cite{ks84} (section II-5).
For the massive neutrinos we use the imperfect fluid treatment
of ref. \cite{schaefer91}.  We numerically integrate the equations
using a Haming predictor-corrector routine as we have found it
tracks the oscillations of the relativistic components more accurately
than Bulirsch-Stoer or Runge-Kutta routines.  After
recombination is completed, ($z=900$) we switch from integrating the
baryon-photon difference equations to simply integrating the baryon and
photon component ($\Delta_{c,baryon},\ \Delta_{cr}$ and $V_{baryon}$, $V_r$)
equations separately.

   We have checked the results of our code by comparing our baryonic transfer
functions against those of ref. \cite{holtzman89} who gives values for
$\Omega_{baryon} = 0.1$ and $0.01$ and found good agreement, although here we
present results only for $\Omega_{baryon}=0.05$.  We have fit the transfer
functions to an inverse 5th order polynomial
and the results are given in table 2.  We find a 5th order inverse polynomial
works a little better for C+HDM models than a fourth order as is more usual
for CDM models.  The transfer functions given here are not baryonic transfer
functions, but rather are fits to the total density
perturbation ({\it i.e.,} $\Delta = \Omega_{CDM} \Delta_{c,CDM} + \Omega_\nu
\Delta_{c\nu} + \Omega_{baryon} \Delta_{c,baryon}$.  We define our transfer
function as
\begin{equation}
T(k) = {\Delta(k, t_0)\over \Delta(k,t_i)}
{\Delta(k=0, t_i)\over \Delta(k=0,t_0)}
\end{equation}
where $t_i$ and $t_0$ are the initial and present times, respectively.
The transfer functions are accurate to a few percent down to $k=1\ h/$Mpc.
\begin{equation}
T(k) = {1\over 1 + t_1 k^{0.5} + t_2 k^1 + t_3 k^{1.5} + t_4 k^2 +t_5 k^{2.5}},
\end{equation}
In table 2, all coefficients are for $k$ in Mpc (with h=0.5).
\section{Estimating the Contribution of Non-linear Effects.}

   In the previous section we alluded to the fact that non-linear effects were
becoming important at $8h^{-1}$ Mpc.  We would like to estimate the size of the
non-linear effects.  To get a crude estimate we use the ``spherical collapse
model" treatment (see, e.g., ref. \cite{peebles80}).  This approximates a
spherical overdensity in a flat universe locally as a miniature closed
collapsing universe, so one
can follow the collapse into the non-linear regime.  We can define the
non-linear overdensity as $\sigma^{\rm non-lin}$ which is given by the
following equation
\begin{equation}
\sigma^{\rm non-lin}= {\rho\over \rho_b}-1 = {9\over 2}{ (\theta - {\rm
sin}\theta)^2\over (1 - {\rm cos}\theta)^3},
\end{equation}
where $\rho_b$ is the background density and $\theta$ is the conformal time
coordinate which parametrizes a closed spacetime.  For the same value of
$\theta$ the linear theory predicts $\sigma^{(\rm lin)}$
\begin{equation}
\sigma^{(\rm lin)} = {3\over 20} \left[ 6(\theta -
{\rm sin}\theta)\right]^{2/3},
\end{equation}
Thus if $\sigma_{\rm (non-lin)} = 0.6$, we can estimate $\sigma^{(\rm lin)} =
0.4$, a 50\% correction.  To keep within a range where the perturbations are
linear we must restrict ourselves to scales where $\sigma \le 0.4$, where
non-linear corrections are estimated to be $\le$ 30\%.

 To estimate what value of $\sigma_m$ the counts in cells at $\ell = 20
h^{-1}$ Mpc implies, we must first correct for redshift space effects
\begin{equation}
\sigma^2(\ell) = b_I^2\left[1 + {2\over 3 b_I} + {1\over 5 b_I^2}\right]
\sigma_m^2(\ell)
\end{equation}
We will assume for now that $b_I = 1.2$, which implies $\sigma_m = 0.4$.  Thus
we should not consider data on scales
${\mathrel{\raise.3ex\hbox{$<$\kern-.75em\lower1ex\hbox{$\sim$}}}} 20 h^{-1}$
Mpc.

\newpage

\begin{figure}
\caption{
  Transfer functions for the cold plus hot dark matter models.  The curves
represent the relative growth of density fluctuations as a function of scale.
The curves represent the present time (z=0) transfer functions and in a
universe with $\Omega_{CDM} + \Omega_{HDM} +\Omega_{baryon}=1$, where we have
taken the canonical value of $\Omega_{baryon}=0.05$ from primordial
nucleosynthesis.}
\label{pwrnu}
\end{figure}

\begin{figure}
\caption{
We show the calculated model power spectra using our best fit model parameters
for $n=1.00$ only.  The values of $b_8$ and $b_I$ have been fit as described in
the text.  The values of $b_I$ are close to 1.0, with the values $b_I=$ 1.1,
1.1, 0.9, 0.9, and 1.0 for $\Omega_{HDM} =$ 0.0, 0.1, 0.2, 0.3, and
0.4, respectively.  }
\label{qdot93}
\end{figure}

\begin{figure}
\caption{
The counts in redshift space cells data from the IRAS QDOT survey [25].
Illustrated are the curves for some selected
models.  Low mass fluctuation amplitudes of some models are compensated by
using high values of the IRAS bias $b_I$ as can be seen here.}
\label{qdcinc}
\end{figure}

\begin{figure}
\caption{
 The values of the bulk streaming velocities extracted by the  POTENT analysis
program with 1 $\sigma$ error bars [29].
Also shown are predicted velocity curves
for selected models.  The models with low $n$ and gravity waves can be ruled
out by this data even though the cosmic variance for the velocity predictions
in a single region are quite large. To illustrate this we have plotted the 68\%
confidence limits on the prediction of large scale streaming velocities for the
25\% HDM model.  }
\label{cphvel}
\end{figure}

\begin{figure}
\caption{
Here we show the IRAS QDOT power spectrum
data [27] 
and the COBE amplitude constraint [23] 
 (converted to a
redshift space power spectrum constraint for $b_I=1.0$).  The models shown have
negligible gravity wave contributions to the COBE anisotropy, and have best fit
normalizations $b_8=1.26$, 1.41, and 1.54 for the models shown
with 0\%, 25\%, and 45\% HDM, respectively.  We have shown only half of the
QDOT data so the plot is easier to read.  The value of $n=.84$ yields the best
fitting CDM model without gravity waves.  Note that simply tilting a CDM
spectrum cannot reproduce the large scale ``bump" in the IRAS power spectrum,
which is generally why CDM models do not fare as well as C+HDM models.
}
\label{pwrcob}
\end{figure}

\begin{figure}
\caption{
Contour plot for $\chi^2$ - No gravity waves
$\chi^2$ contours in the $n$-$\Omega_{HDM}$ plane for models with a negligible
gravity wave content.  Moving outward from the center of the graph are contours
corresponding to .5 \% (dotted), 1\% (dash-dotted), 5\% (dashed), 10\% (heavy
dashed) 25\% (long dashed), 50\% (heavy long dashed), 68\% (solid), and 95\%
(heavy solid) confidence levels.
}
\label{l}
\end{figure}

\begin{figure}
\caption{
Contour plot for $\chi^2$ - With gravity waves
$\chi^2$ contours in the $n$-$\Omega_{HDM}$ plane for models with
gravity waves. Moving outward from the center of the graph are contours
corresponding to .5 \% (dotted), 1\% (dash-dotted), 5\% (dashed), 10\% (heavy
dashed) 25\% (long dashed), 50\% (heavy long dashed), 68\% (solid), 95\%
(solid), 99\% (heavy solid) confidence levels.
}
\label{lgw}
\end{figure}

\begin{figure}
\caption{
Same as figure 6, but we have added the constraints from non-linear data.  We
have added the restriction that $b_8\geq 1.25$ and that we have sufficient
power for quasar formation.  The procedure for adding these constraints in a
$\chi^2$ analysis is described in the text.  The significance of the
contours is here not well defined because of the way we have added the
non-linear constraints.
}
\label{nl}
\end{figure}

\begin{figure}
\caption{
Same as figure 7, but we have added the constraints from non-linear data.  We
have added the restriction that $b_8\geq 1.25$ and that we have sufficient
power for quasar formation.  The significance of the contours is here not
precise because of the way we have added the non-linear constraints.
}
\label{nlgw}
\end{figure}

\begin{table}
\caption{$k_{eff}$ probed by counts in cells of volume $\ell^3$}
\label{table1}
\begin{tabular}{| l  r |}
$\ell$      & $k_{eff}$     \\ \hline
20 Mpc/$h$  & 0.120 $h/$Mpc \\
30 Mpc/$h$  & 0.081 $h/$Mpc \\
40 Mpc/$h$  & 0.031 $h/$Mpc \\
60 Mpc/$h$  & 0.022 $h/$Mpc
\end{tabular}
\end{table}

\begin{table}
\label{table2}
\caption{transfer functions with $\Omega_{baryon}=0.05$, $h=0.5$}
\begin{tabular}{|l|r|r|r|r|r|}
  $\Omega_\nu$ & $t_1$   & $t_2$  &  $t_3$  & $t_4$   & $t_5$  \\ \hline\hline
   0.00 & -1.150  &  29.60 & 48.49 & -43.17 &  132.4 \\ \hline
   0.05 & -0.8654 &  17.65 & 165.1 & -277.1 &  343.4 \\ \hline
   0.10 & -0.2942 &  1.393 & 274.6 & -472.6 &  538.2 \\ \hline
   0.15 &  0.1157 & -8.820 & 330.0 & -541.4 &  660.8 \\ \hline
   0.20 &  0.3176 & -12.69 & 334.9 & -495.8 &  726.1 \\ \hline
   0.25 &  0.3128 & -10.60 & 296.3 & -361.6 &  756.5 \\ \hline
   0.30 &  0.1363 & -3.540 & 219.1 & -142.9 &  771.2 \\ \hline
   0.35 & -0.1454 &  6.144 & 127.6 &  78.64 &  840.3 \\ \hline
   0.40 & -0.4276 &  15.15 & 53.35 &  193.1 &  1084. \\ \hline
   0.45 & -0.6522 &  21.44 & 15.98 &  131.2 &  1582. \\ \hline
   0.50 & -0.7882 &  24.04 & 25.49 & -147.3 &  2395.
\end{tabular}
\end{table}

\end{document}